\documentclass{article}%
\usepackage{amsfonts}
\usepackage{amsmath}
\usepackage{amssymb}
\usepackage{graphicx}
\usepackage[margin=2cm]{geometry}
\usepackage{hyperref}%
\providecommand{\U}[1]{\protect\rule{.1in}{.1in}}
\newtheorem{theorem}{Theorem}
\newtheorem{acknowledgement}[theorem]{Acknowledgement}

\begin{document}

\title{Cosmological Dark Matter Amplification through Dark Torsion}
\author{Fernando Izaurieta$^{1}$ and Samuel Lepe$^{2}$\\$^{1}$\textit{Departamento de F\'{\i}sica, Universidad de Concepci\'{o}n},\\Casilla 160-C, Concepci\'{o}n, Chile,\\\texttt{fizaurie@udec.cl}\\$^{2}$\textit{Instituto de F\'{\i}sica, Pontificia Universidad Cat\'{o}lica de
Valpara\'{\i}so},\\Av. Brasil 2950, Valpara\'{\i}so, Chile,\\\texttt{samuel.lepe@pucv.cl}}
\maketitle

\begin{abstract}
A cosmological approach based on considering a cosmic background with non-zero
torsion is shown in order to give an option of explaining a possible phantom
evolution, not ruled out according to the current observational data. We
revise some aspects of the formal schemes on torsion and, according them, we
develop a formalism which can be an interesting alternative for exploring Cosmology.

\end{abstract}

\section*{Introduction}

The current knowledge of the nature of dark matter is scarce. However, the
cumulative evidence seems to favor the scenario of dark matter as a
non-interacting form of matter instead of some modified gravity theory. This
is true in particular when considering phenomena as cluster collisions (e.g.
Refs.~\cite{Ref-DM-Bullet,Ref-DM-Cluster2}).

Our ignorance of dark matter physical behavior implies a lack of knowledge of
its features as a source of gravity. In particular, it is unknown whether or
not the spin tensor of dark matter vanishes. The issue is relevant since a
non-vanishing spin tensor is a source of torsion, and torsion requires to go
beyond General Relativity (GR). The closest framework to General Relativity is
the Einstein-Cartan-Sciama-Kibble (ECSK) theory of gravity. There are many
other alternatives theories of gravity with torsion, but ECSK is probably the
simplest one that includes spinning matter and torsion.

The ignorance regarding the physical nature of dark matter is in sharp
contrast with the knowledge on the behavior of the Standard Model matter. For
instance, from the Yang-Mills (YM) Lagrangian%

\begin{equation}
\mathcal{L}_{\mathrm{YM}}=-\frac{1}{4}F^{A}{}_{\mu\nu}F^{B\mu\nu}\,
\mathrm{tr}\left(  \boldsymbol{T}_{A}\boldsymbol{T}_{B}\right)  ,
\end{equation}
it is straightforward to see that the spin tensor of Yang-Mills bosons
vanishes. Therefore, Yang-Mills bosons are not a source of torsion, and there
is no Yang-Mills bosons-torsion interaction. In contrast, the Standard Model
fermions have a non-vanishing spin tensor, and therefore they are a source of
torsion. They should interact with torsion, but the effect is so feeble that
it is hard to foresee any particle-physics experiment capable of detecting it
(See Chap. 8.4 of Ref.~\cite{Ref-SUGRA-Van-Proeyen}).

Torsion does not interact with matter at a classical level (see
Ref.~\cite{Ref-Hehl-GravProbeB}), and neither do with electromagnetic
phenomena. For instance, regardless of background torsion, classical point
particles should follow torsionless geodesics, and electromagnetic waves
should travel trough torsionless null geodesics. To Standard Model matter,
torsion is \textquotedblleft dark.\textquotedblright\ Perhaps, the only
realistic way of detecting torsion could be through precise measurement of the
polarization of gravitational waves (see
Refs.~\cite{Ref-Nos-2019-GW-Polarization} and~\cite{Ref-Nos-2019-GW-Torsion}).

Even further, due to interaction and decoherence, Standard Model baryons are
highly localized, and they form astrophysical structures. In the context of
ECSK theory, torsion is not able to propagate in a vacuum (in glaring contrast
to the behavior of Riemannian curvature). Therefore, given both the granular
nature of the baryonic matter in the current epoch of the Universe and that
torsion vanishes in the vacuum, it seems incorrect to associate an effective
non-vanishing spin tensor to Standard Model matter in cosmological scales in
modern times. In other words, it seems unrealistic (see
Ref.~\cite{Ref-Anti-Weyssenhof}) to consider Standard Model baryons as a spin
fluid on a cosmological scale: the effective spin tensor of a gas of galaxies
vanishes in long scales.

In contrast, the spin tensor of Standard Model matter is a relevant source of
torsion in a Universe filled with a high-density plasma of Standard Model
fermions. That is the case of bounce models at the very early Universe (see
Ref.~\cite{Ref-Poplawski-Big-Bounce}). In this model, the torsion created by
high-density fermion plasma gives rise to inflation-like behaviors at very
early times.

The situation is arguably different for dark matter. Its lack of interaction
with Standard Model matter and its incapability to create dark matter
structures lead to the conjecture that the decoherence effects could be feeble
for dark matter. Even more, this picture of dark matter fits well with its
distribution being broader and more unlocalized than the one of Standard Model
matter. Therefore, if dark matter has a non-vanishing spin tensor, it seems
natural to expect that it could give rise to torsion in cosmological scales.

The torsion created through this mechanism would be as dark as its source:
Standard Model matter would not be able to interact with it. When moving these
\textquotedblleft dark torsion\textquotedblright\ terms to the right-hand side
of the field equations, they behave just as an extra (and dark) source of
standard torsionless Riemannian curvature.

From an observational point of view, it is possible to measure only the
Riemannian curvature and not the torsion. Therefore, in this scenario, the
observed gravitational dark matter effects correspond to the ones created by
the "bare" dark matter plus the torsional \textquotedblleft dark
dress\textquotedblright\ it creates through its spin tensor.

The current article explores the idea of how \textquotedblleft dark
torsion\textquotedblright\ could amplify the effects of a small amount of dark
matter in a cosmological setting. Given the disparity between the amount of
dark matter and Standard Model baryons in the Universe, a mechanism as this
one may seem of interest. The Section~1 briefly reviews ECSK gravity and shows
how torsion amplifies the effects of \textquotedblleft bare dark
matter\textquotedblright,\ creating a higher effective energy density. This
total torsion-dressed density would correspond to the observed dark matter
density instead of the bare piece. In the case of Standard Model fermions, the
canonical approach is to describe their spin tensor as a Weyssenhof fluid.
However, given the lack of dark matter self-interaction, this Ansatz does not
seem correct. In this Section, we offer a different Ansatz for the spin tensor
of dark matter using symmetry and dimensional analysis arguments. The
Sections~1.1 and 2 uses the generalized Friedmann equations to analyze the
cosmological consequences of torsion and its dark matter amplification effect.
The Section~3 studies the thermodynamical effects of the torsional dress of
dark matter. Finally, in Section~\ref{Sec_TheEnd} we present some conclusions
and possible further works.

\label{Sec_DM-DT}

\section{Dark Matter and Dark Torsion}

There are many works in the context of cosmology using alternative theories of
gravity involving a non-vanishing torsion
(\cite{Ref-Poplawski-Big-Bounce,Ref-Pasmatsiou,Ref-Kranas,Ref-Cabral,Ref-Magueijo,Ref-Alexander,Ref-Nos-2018-CosmoHorndsk}%
). The present work focuses on the most straightforward approach, i.e., ECSK
theory. It also the closest to standard GR. The idea is to have a taste of
some of the consequences of non-vanishing spin tensor for dark matter in the
simplest context before considering more exotic approaches.

Let us consider a four-dimensional spacetime with $\left(  -,+,+,+\right)  $
signature described by the Einstein--Cartan geometry, i.e., the metric
$g_{\mu\nu}$ and the connection $\Gamma_{\mu\nu}^{\lambda}$ are independent
degrees of freedom. The ECSK action principle corresponds to
\begin{equation}
\mathcal{S}=\int\sqrt{\left\vert g\right\vert }\mathrm{d}^{4}x\left(
\mathcal{L}_{\mathrm{G}}+\mathcal{L}_{\mathrm{b}}+\mathcal{L}_{\mathrm{DM}%
}\right)  , \label{Eq_Action}%
\end{equation}
where we are using units $c=8\pi G=k_{\mathrm{B}}=1$. In Eq.~(\ref{Eq_Action})
$\mathcal{L}_{\mathrm{b}}$ stands for the Lagrangian for baryonic matter and
$\mathcal{L}_{\mathrm{DM}}$ corresponds to an unknown Lagrangian for dark
matter. The gravity Lagrangian $\mathcal{L}_{\mathrm{G}}$ corresponds to the
standard Einstein--Hilbert term a la Palatini, i.e., without imposing the
torsionless condition (and therefore with the metric and the connection as
independent degrees of freedom),%
\begin{equation}
\mathcal{L}_{\mathrm{G}}\left(  g,\Gamma,\partial\Gamma\right)  =\frac{1}%
{2}R\left(  g,\Gamma,\partial\Gamma\right)  -\Lambda.
\end{equation}
Here $R=g^{\sigma\nu}R^{\mu}{}_{\sigma\mu\nu}$ is the generalization of the
Ricci scalar constructed from the generalized Riemann tensor (or Lorentz curvature)%

\begin{equation}
R^{\rho}{}_{\sigma\mu\nu}=\partial_{\mu}\Gamma_{\nu\sigma}^{\rho}%
-\partial_{\nu}\Gamma_{\mu\sigma}^{\rho}+\Gamma_{\mu\lambda}^{\rho}\Gamma
_{\nu\sigma}^{\lambda}-\Gamma_{\nu\lambda}^{\rho}\Gamma_{\mu\sigma}^{\lambda},
\end{equation}
where $\Gamma_{\nu\sigma}^{\rho}$ is a general connection (not necessarily the
Christoffel one).

The action principle, Eq.~(\ref{Eq_Action}), may seem general. However, it is
fair to remark that the Lagrangian choice Eq.~(\ref{Eq_Action}) assumes the
minimal coupling between dark matter, baryons, and gravity, and it does not
include torsional terms as the Holst term. Non-minimal couplings with
gravitational terms are sources of torsion, even for scalar bosonic
fields~\cite{Ref-Nos-2017-Horndeski} and in cosmological
settings~\cite{Ref-Nos-2018-CosmoHorndsk}. Similarly, non-minimal couplings
within the Standard Model piece give rise to axions, which may be a promising
dark matter candidate. Therefore it is worth remembering that
Eq.~(\ref{Eq_Action}) corresponds to the simplest ECSK case and there are many
other more exotic choices.

The antisymmetric part of the connection $\Gamma_{\mu\nu}^{\lambda}$ defines
the torsion tensor as%
\begin{equation}
T^{\lambda}{}_{\mu\nu}=\Gamma_{\mu\nu}^{\lambda}-\Gamma_{\nu\mu}^{\lambda},
\end{equation}
and the difference between the general connection $\Gamma_{\mu\nu}^{\lambda}$
and the canonical Christoffel connection $\mathring{\Gamma}_{\mu\nu}^{\lambda
}=\left(  1/2\right)  g^{\lambda\rho}\left(  \partial_{\mu}g_{\nu\rho
}+\partial_{\nu}g_{\mu\rho}-\partial_{\rho}g_{\mu\nu}\right)  $ is given by%
\begin{equation}
\Gamma_{\mu\nu}^{\lambda}-\mathring{\Gamma}_{\mu\nu}^{\lambda}=K^{\lambda}%
{}_{\nu\mu},
\end{equation}
where the right-hand side corresponds to the contorsion\footnote{It seems
there is no agreement in the literature on the name of this tensor. Some
authors call it \textquotedblleft contortion,\textquotedblright\ while others
use \textquotedblleft contorsion.\textquotedblright\ We have chosen to use the
later one because it sounds closer to torsion. The word \textquotedblleft
contortion\textquotedblright\ may also be confused with a twisting motion.}
tensor
\begin{equation}
K_{\mu\nu\lambda}=\frac{1}{2}\left(  T_{\nu\mu\lambda}-T_{\mu\nu\lambda
}+T_{\lambda\mu\nu}\right)  . \label{Eq_K=T}%
\end{equation}
It is possible to decompose the generalized curvature in terms of the
contorsion as%
\begin{equation}
R^{\alpha\beta}{}_{\mu\nu}=\mathring{R}^{\alpha\beta}{}_{\mu\nu}%
+\mathring{\nabla}_{\mu}K^{\alpha\beta}{}_{\nu}-\mathring{\nabla}_{\nu
}K^{\alpha\beta}{}_{\mu}+K^{\alpha}{}_{\lambda\mu}K^{\lambda\beta}{}_{\nu
}-K^{\alpha}{}_{\lambda\nu}K^{\lambda\beta}{}_{\mu}, \label{Eq_R+DK+K2}%
\end{equation}
where $\mathring{R}^{\alpha\beta}{}_{\mu\nu}$ is the canonical torsionless
Riemann tensor in terms of the Christoffel connection $\mathring{\Gamma}%
_{\mu\nu}^{\lambda}$, and $\mathring{\nabla}_{\mu}$ is the standard
torsionless covariant derivative in terms of it.

The metric equations of motion are given by%
\begin{equation}
R_{\mu\nu}^{+}-\frac{1}{2}g_{\mu\nu}R+\Lambda g_{\mu\nu}=\mathcal{T}_{\mu\nu
}^{\left(  \mathrm{b}\right)  }+\mathcal{T}_{\mu\nu}^{\left(  \mathrm{DM}%
\right)  }, \label{Eq_Field_Metric}%
\end{equation}
where $R_{\mu\nu}^{+}$ is the symmetric part of the generalized Ricci
tensor\footnote{In the case of non-vanishing torsion, the generalized Ricci
tensor has an antisymmetric part given by $R_{\mu\nu}^{-}=\frac{1}{2}\left(
R_{\mu\nu}-R_{\nu\mu}\right)  =-\frac{1}{2}\left(  \nabla_{\lambda}T^{\lambda
}{}_{\mu\nu}+\nabla_{\nu}T^{\lambda}{}_{\lambda\mu}-\nabla_{\mu}T^{\lambda}%
{}_{\lambda\nu}+T^{\lambda}{}_{\rho\lambda}T^{\rho}{}_{\mu\nu}+T^{\rho}%
{}_{\lambda\mu}T^{\lambda}{}_{\rho\nu}-T^{\rho}{}_{\lambda\nu}T^{\lambda}%
{}_{\rho\mu}\right)  .$} and $\mathcal{T}_{\mu\nu}^{\left(  \mathrm{b}\right)
}$ and $\mathcal{T}_{\mu\nu}^{\left(  \mathrm{DM}\right)  }$ are the
stress-energy tensors associated with $\mathcal{L}_{\mathrm{b}}$ and
$\mathcal{L}_{\mathrm{DM}}$. The affine equations of motion are given by%
\begin{equation}
T_{\lambda\mu\nu}-g_{\lambda\mu}T^{\rho}{}_{\rho\nu}+g_{\lambda\nu}T^{\rho}%
{}_{\rho\mu}=\sigma_{\lambda\mu\nu}^{\left(  \mathrm{b}\right)  }%
+\sigma_{\lambda\mu\nu}^{\left(  \mathrm{DM}\right)  },
\label{Eq_Field_Affine}%
\end{equation}
where $\sigma_{\lambda\mu\nu}^{\left(  \mathrm{b}\right)  }=-\sigma
_{\lambda\nu\mu}^{\left(  \mathrm{b}\right)  }$ and $\sigma_{\lambda\mu\nu
}^{\left(  \mathrm{DM}\right)  }=-\sigma_{\lambda\nu\mu}^{\left(
\mathrm{DM}\right)  }$ are the spin tensors\footnote{The spin tensor is the
variation of the matter Lagrangian with respect to the connection, in the same
way as the stress-energy tensor is the variation of the matter Lagrangian with
respect to the metric. The spin tensor of classical matter (e.g., dust)
vanishes, but the spin tensor of a fermionic particle does not. For instance,
the spin tensor of an electron is proportional to its axial current.}
associated with $\mathcal{L}_{\mathrm{b}}$ and $\mathcal{L}_{\mathrm{DM}}$.

We would like to end this brief review of ECSK pointing that in general
$\nabla^{\mu}\left(  R_{\mu\nu}-\frac{1}{2}g_{\mu\nu}R\right)  \neq0$ and
therefore the right-hand side of Eq.~(\ref{Eq_Field_Metric}) is not longer
\textquotedblleft conserved\textquotedblright. It is possible to write down a
genuine conservation law using Eq.~(\ref{Eq_R+DK+K2}) to move all the
torsional terms to the right-hand side%
\begin{equation}
\mathring{R}_{\mu\nu}-\frac{1}{2}g_{\mu\nu}\mathring{R}+\Lambda g_{\mu\nu
}=\mathcal{T}_{\mu\nu}^{\left(  \mathrm{b}\right)  }+\mathcal{T}_{\mu\nu
}^{\left(  \mathrm{DM}\right)  }+\mathcal{T}_{\mu\nu}^{\left(  \mathrm{T}%
\right)  }, \label{Eq_TorsionRightHandSide}%
\end{equation}
where $\mathring{R}_{\mu\nu}$ is the standard torsionless Ricci tensor and
$\mathcal{T}_{\mu\nu}^{\left(  \mathrm{T}\right)  }$ is the effective
stress-energy tensor for torsion given by%
\begin{align}
\mathcal{T}_{\mu\nu}^{\left(  \mathrm{T}\right)  }  &  =g_{\mu\nu}\left(
\mathring{\nabla}_{\alpha}K^{\alpha\rho}{}_{\rho}+\frac{1}{2}\left[
K^{\alpha}{}_{\lambda\alpha}K^{\lambda\rho}{}_{\rho}-K^{\alpha}{}_{\lambda
\rho}K^{\lambda\rho}{}_{\alpha}\right]  \right)  +\nonumber\\
&  +\frac{1}{2}\left(  \mathring{\nabla}_{\nu}K^{\alpha}{}_{\mu\alpha
}+\mathring{\nabla}_{\mu}K^{\alpha}{}_{\nu\alpha}+K^{\alpha}{}_{\lambda\mu
}K^{\lambda}{}_{\nu\alpha}+K^{\alpha}{}_{\lambda\nu}K^{\lambda}{}_{\mu\alpha
}-\left[  \mathring{\nabla}_{\lambda}+K^{\alpha}{}_{\lambda\alpha}\right]
\left[  K^{\lambda}{}_{\mu\nu}+K^{\lambda}{}_{\nu\mu}\right]  \right)  .
\label{Eq_T_eff_torsion}%
\end{align}
Doing this, the \textquotedblleft conservation law\textquotedblright\ takes
the form%
\begin{equation}
\mathring{\nabla}^{\mu}\left(  \mathcal{T}_{\mu\nu}^{\left(  \mathrm{b}%
\right)  }+\mathcal{T}_{\mu\nu}^{\left(  \mathrm{DM}\right)  }+\mathcal{T}%
_{\mu\nu}^{\left(  \mathrm{T}\right)  }\right)  =0.
\end{equation}
As mentioned in the Introduction, the baryons spin tensor $\sigma_{\lambda
\mu\nu}^{\left(  \mathrm{b}\right)  }$ and the torsion associated with it
could have been relevant under the extremely high fermion densities of the
very early Universe~\cite{Ref-Poplawski-Big-Bounce}. However, in current times
$\sigma_{\lambda\mu\nu}^{\left(  \mathrm{b}\right)  }=0$ should be an
excellent approximation for any cosmological purpose.

Considering $\sigma_{\lambda\mu\nu}^{\left(  \mathrm{b}\right)  }=0$ and
tracing the affine equation of motion~(\ref{Eq_Field_Affine}), it is clear
that in cosmological scales we should have%
\begin{equation}
T_{\lambda\mu\nu}=\sigma_{\lambda\mu\nu}^{\left(  \mathrm{DM}\right)  }%
+\frac{1}{2}\left[  g_{\lambda\nu}\sigma^{\rho}{}_{\rho\mu}^{\left(
\mathrm{DM}\right)  }-g_{\lambda\mu}\sigma^{\rho}{}_{\rho\nu}^{\left(
\mathrm{DM}\right)  }\right]  , \label{Eq_T=DM}%
\end{equation}
which means that torsion vanishes in the absence of dark matter. In the
context of ECSK, torsion cannot propagate in a vacuum. To have a propagating
torsion, we must have a different action choice than Eq.~(\ref{Eq_Action}),
for instance, the Holst action or the Horndeski generalization of
Refs.~\cite{Ref-Nos-2017-Horndeski,Ref-Nos-2018-CosmoHorndsk}.

Since torsion is dark for baryonic matter, Eq.~(\ref{Eq_Field_Metric}) can be
regarded as%
\begin{equation}
\mathring{R}_{\mu\nu}-\frac{1}{2}g_{\mu\nu}\mathring{R}+\Lambda g_{\mu\nu
}=\mathcal{T}_{\mu\nu}^{\left(  \mathrm{b}\right)  }+\mathcal{T}_{\mu\nu
}^{\left(  \mathrm{eff-DM}\right)  },
\end{equation}
where the effective dark matter stress-energy tensor $\mathcal{T}_{\mu\nu
}^{\left(  \mathrm{eff-DM}\right)  }=\mathcal{T}_{\mu\nu}^{\left(
\mathrm{DM}\right)  }+\mathcal{T}_{\mu\nu}^{\left(  \mathrm{T}\right)  }$
causes the observed effects of dark matter. Here $\mathcal{T}_{\mu\nu
}^{\left(  \mathrm{DM}\right)  }$ corresponds to the stress-energy tensor of
\textquotedblleft bare\textquotedblright\ dark matter. In the next sections,
we show that torsion amplifies its weight through the effective torsional
stress-energy tensor $\mathcal{T}_{\mu\nu}^{\left(  \mathrm{T}\right)  }$ from
Eq.~(\ref{Eq_T_eff_torsion}). Since the interaction of torsion with baryonic
matter is negligible in current times, torsion would be as dark as its source.
Since current observations are only able to detect the Riemannian gravity
piece of the geometry, they would be sensitive to the combined or
\textquotedblleft dressed\textquotedblright\ effect of $\mathcal{T}_{\mu\nu
}^{\left(  \mathrm{eff-DM}\right)  }=\mathcal{T}_{\mu\nu}^{\left(
\mathrm{DM}\right)  }+\mathcal{T}_{\mu\nu}^{\left(  \mathrm{T}\right)  }$, but
they won't be able to distinguish bare dark matter from torsion. Perhaps, only
a careful measurement of the propagation of polarization of gravitational
waves could distinguish bare dark matter from its \textquotedblleft torsional
dress\textquotedblright~\cite{Ref-Nos-2019-GW-Polarization}.

At this point, the lack of precise knowledge of the nature of dark matter
creates what may seem like an insurmountable problem when trying to model its
spin tensor. There are some usual Ans\"{a}tze for the spin tensor, as the
Weyssenhof fluid~\cite{Ref-Weyssenhof, Ref-Obukhov-Weyssenhof}. However, given
our ignorance on the physics of dark matter, any Ansatz for $\sigma
_{\lambda\mu\nu}^{\left(  \mathrm{DM}\right)  }$ may seem excessive. The
problem is that since we do not have any information on the spin tensor of
dark matter $\sigma_{\lambda\mu\nu}^{\left(  \mathrm{DM}\right)  }$, we cannot
use the field equation~(\ref{Eq_T=DM}). Without this field equation we do not
have information on torsion and it seems impossible to solve the system.

In what follows this problem is treated in a cosmological setting. The general
idea is that whatever dark matter is, it is possible to use symmetry arguments
and dimensional analysis to arrive at a general Ansatz for an effective
$\sigma_{\lambda\mu\nu}^{\left(  \mathrm{DM}\right)  }$ in cosmological
scales. Using this Ansatz, it becomes possible to study the effects of the
torsion created by dark matter in cosmic evolution.

Let us start by considering the canonical FLRW metric with a homogeneous,
isotropic and Riemannian flat spatial section%
\begin{equation}
\mathrm{d}s^{2}=-\mathrm{d}t^{2}+a^{2}\left(  t\right)  \left(  \mathrm{d}%
x^{2}+\mathrm{d}y^{2}+\mathrm{d}z^{2}\right)  . \label{Eq_FLRW_Metric}%
\end{equation}
Despite not knowing the Lagrangian $\mathcal{L}_{\mathrm{DM}}$, we know that
the only stress-energy tensor compatible with the cosmological symmetries
$\pounds _{\zeta}\mathcal{T}_{\mu\nu}^{\left(  \mathrm{DM}\right)  }=0$ is the
canonical%
\begin{equation}
\mathcal{T}_{\mu\nu}^{\left(  \mathrm{DM}\right)  }=\left(  \rho_{\mathrm{DM}%
}+p_{\mathrm{DM}}\right)  U_{\mu}U_{\nu}+p_{\mathrm{DM}}g_{\mu\nu},
\end{equation}
where $\rho_{\mathrm{DM}}$ and $p_{\mathrm{DM}}$ are the dark matter density
and pressure. With the spin tensor it is possible to do the same. The most
general spatially isotropic and homogeneous spin tensor $\pounds _{\zeta
}\sigma_{\lambda\mu\nu}^{\left(  \mathrm{DM}\right)  }=0$ for dark matter must
have the \textquotedblleft Cartan staircase\textquotedblright\ form%
\begin{equation}
\sigma_{\lambda\mu\nu}^{\left(  \mathrm{DM}\right)  }=-2\left(  g_{\mu\lambda
}g_{\nu\rho}-g_{\mu\rho}g_{\nu\lambda}\right)  h^{\rho}\left(  t\right)
-2\sqrt{\left\vert g\right\vert }\epsilon_{\lambda\mu\nu\rho}f^{\rho}\left(
t\right)  , \label{Ec_Spin_Tensor_Escalera_Cartan}%
\end{equation}
with the 4-vectors $h^{\rho}\left(  t\right)  $ and $f^{\rho}\left(  t\right)
$ having the form $h^{\rho}\left(  t\right)  =-h\left(  t\right)  U^{\rho}$
and $f^{\rho}\left(  t\right)  =-f\left(  t\right)  U^{\rho}$. In terms of
components%
\begin{align}
\sigma_{0\mu\nu}^{\left(  \mathrm{DM}\right)  }  &
=0,\label{Ec_Spin_Tensor-0}\\
\sigma_{ij0}^{\left(  \mathrm{DM}\right)  }  &  =2g_{ij}h\left(  t\right)
,\label{Ec_Spin_Tensor-h}\\
\sigma_{ijk}^{\left(  \mathrm{DM}\right)  }  &  =2\sqrt{\left\vert
g\right\vert }\epsilon_{ijk}f\left(  t\right)  , \label{Ec_Spin_Tensor-f}%
\end{align}
where $i,j,k,=1,2,3$ and $\lambda,\mu,\nu,=0,1,2,3$. From
Eqs.~(\ref{Ec_Spin_Tensor-0}-\ref{Ec_Spin_Tensor-f}) it is already clear that
dark matter spatial isotropy and homogeneity are not fully compatible with
usual models as the Weyssenhoff spin fluid. At this point, we start to notice
how different it is to model a high-density fermionic plasma as a source of
spin and torsion and non-interacting dark matter. A high-density fermionic
plasma in the early universe is well modeled as a Weyssenhoff spin fluid
because their strong interactions create a rapidly changing spin tensor in
short scales. It respects the Copernican principle because in longer
cosmological scales only matters the average of these local spin anisotropies.
The same arguments do not seem to hold for dark matter, considering it as a
non-interacting fluid extended over cosmological distances in the current
epoch. That is why the Ansatz Eq.~(\ref{Ec_Spin_Tensor_Escalera_Cartan}) for
the spin tensor of dark matter may be a far better choice than the standard
Weyssenhoff spin fluid.

The spin tensor, torsion and contorsion are all algebraically related through
equations~(\ref{Eq_T=DM}) and~(\ref{Eq_K=T}). From them, it is straightforward
to conclude that whatever $h\left(  t\right)  $ and $f\left(  t\right)  $ are%
\begin{align}
T_{\lambda\mu\nu}  &  =\left(  g_{\mu\lambda}g_{\nu\rho}-g_{\mu\rho}%
g_{\nu\lambda}\right)  h^{\rho}\left(  t\right)  -2\sqrt{\left\vert
g\right\vert }\epsilon_{\lambda\mu\nu\rho}f^{\rho}\left(  t\right)
,\label{Eq_FLRW_T}\\
K_{\mu\nu\lambda}  &  =\left(  g_{\mu\lambda}g_{\nu\rho}-g_{\mu\rho}%
g_{\nu\lambda}\right)  h^{\rho}\left(  t\right)  +\sqrt{\left\vert
g\right\vert }f^{\rho}\left(  t\right)  \epsilon_{\rho\mu\nu\lambda},
\label{Eq_FLRW_K}%
\end{align}
i.e., the same (still unknown) functions $h\left(  t\right)  $ and $f\left(
t\right)  $ describe torsion and contorsion.

Using the expressions~(\ref{Eq_FLRW_Metric},\ref{Eq_FLRW_T},\ref{Eq_FLRW_K}),
it is possible to calculate the Lorentz curvature components~(\ref{Eq_R+DK+K2}%
) and from it the field equations~(\ref{Eq_Field_Metric}) lead to the
generalized Friedmann relations%
\begin{align}
3\left[  \left(  H+h\right)  ^{2}-f^{2}\right]   &  =\rho_{\mathrm{DM}%
},\label{Eq_Friedmann_Gen_density}\\
2\left(  \dot{H}+\dot{h}\right)  +\left(  3H+h\right)  \left(  H+h\right)
-f^{2}  &  =-p_{\mathrm{DM}}. \label{Eq_Friedmann_Gen_p}%
\end{align}
To solve the Eqs.~(\ref{Eq_Friedmann_Gen_density})
and~(\ref{Eq_Friedmann_Gen_p}), we need to know the dependence of $f$ and $h$
on other physical variables, like dark matter density and pressure. The next
section shows how to find an Ansatz for these \textquotedblleft torsional
equations of state,\textquotedblright\ and to solve the system.

\subsection{Torsional dressing of Dark Matter and Dark Energy}

At this point, instead of making some standard conjecture (Weyssenhoff fluid,
Frenkel condition, Tulczyjew condition, etc.) on the physical nature of the
dark matter spin tensor $\sigma_{\lambda\mu\nu}^{\left(  \mathrm{DM}\right)
}$, we adopted a simpler approach. We may not have an understanding of
$\sigma_{\lambda\mu\nu}^{\left(  \mathrm{DM}\right)  }$ from first principles,
but we have some clues about its form. On the one hand, replacing
Eq.~(\ref{Eq_FLRW_K}) in Eq.~(\ref{Eq_T_eff_torsion}) we can get an effective
stress-energy tensor $\mathcal{T}_{\mu\nu}^{\left(  \mathrm{T}\right)  }$ in
terms of $f$ and $h$. Using dimensional analysis on it, it is clear that at
least in what concerns units we have%
\begin{align}
f  &  \sim\sqrt{\mathrm{energy}\text{ }\mathrm{density}},\\
h  &  \sim\sqrt{\mathrm{energy}\text{ }\mathrm{density}},\\
\mathring{\nabla}_{\mu}h^{\mu}  &  \sim\mathrm{energy}\text{ }\mathrm{density}%
.
\end{align}
On the other hand, it is clear that in a dark matter vacuum ($\rho
_{\mathrm{DM}}=0$) its spin tensor vanishes and $h\left(  t\right)  =f\left(
t\right)  =0$. Similarly, it seems reasonable to expect $h\left(  t\right)  $
and $f\left(  t\right)  $ to grow for higher values of $\rho_{\mathrm{DM}}$.
For this reason, it seems natural to propose an Ans\"{a}tze of
\textquotedblleft barotropic relations\textquotedblright\ between $h\left(
t\right)  $ and $f\left(  t\right)  $ and the dark matter energy density
$\rho_{\mathrm{DM}}$ of the form%
\begin{align}
f  &  \sim\sqrt{\rho_{\mathrm{DM}}},\label{Eq_Protobaro_f}\\
h  &  \sim\sqrt{\rho_{\mathrm{DM}}}. \label{Eq_Protobaro_Dh}%
\end{align}
Of course, much more complex relationships are possible, but these seem to be
the simplest torsional equations of state. Let us consider a standard
barotropic relation for the dark matter pressure $p_{\mathrm{DM}}%
=\omega_{\mathrm{DM}}\rho_{\mathrm{DM}}$ and let us write the barotropic
Ansatz\ for $f$ as%
\begin{equation}
f=\alpha_{f}\sqrt{\frac{\rho_{\mathrm{DM}}}{3}},
\end{equation}
where $\alpha_{f}$ is a constant. In terms of $\alpha_{f}$ it proves practical
to define the \textquotedblleft semi-dressed\textquotedblright\ dark matter
energy density and pressure%
\begin{align}
\rho_{f}  &  =\rho_{\mathrm{DM}}+3f^{2}=\left(  1+\alpha_{f}^{2}\right)
\rho_{\mathrm{DM}},\label{Eq_Semidressed_Density}\\
p_{f}  &  =p_{\mathrm{DM}}-f^{2}=\left(  \omega_{\mathrm{DM}}-\frac{1}%
{3}\alpha_{f}^{2}\right)  \rho_{\mathrm{DM}}. \label{Eq_Semidressed_Pressure}%
\end{align}
In terms of $\rho_{f}$ and $p_{f}$, the Eqs.~(\ref{Eq_Friedmann_Gen_density})
and~(\ref{Eq_Friedmann_Gen_p}) take the simpler form%
\begin{align}
3\left(  H+h\right)  ^{2}  &  =\rho_{f},
\label{Eq_Friedmann_Density_Seminaked}\\
2\left(  \dot{H}+\dot{h}\right)  +\left(  3H+h\right)  \left(  H+h\right)   &
=-p_{f}, \label{Eq_Friedmann_Pressure_Seminaked}%
\end{align}
where the \textquotedblleft semi-dressed\textquotedblright\ pressure $p_{f}$
obeys the barotropic relation $p_{f}=\omega_{f}\rho_{f}$ and
\begin{equation}
\omega_{f}=\frac{\omega_{\mathrm{DM}}-\alpha_{f}^{2}/3}{1+\alpha_{f}^{2}}.
\end{equation}
In short, the $f$-component of the spin tensor has the effect of replacing the
original \textquotedblleft bare\textquotedblright\ dark matter density
$\rho_{\mathrm{DM}}$ by an amplified \textquotedblleft
semi-dressed\textquotedblright\ energy density $\rho_{f}$,
Eq.~(\ref{Eq_Semidressed_Density}). The pressure $p_{\mathrm{DM}}$ is replaced
by an smaller \textquotedblleft semi-dressed\textquotedblright\ $p_{f}$
pressure, Eq.~(\ref{Eq_Semidressed_Pressure}). It is worth to notice that in
the case of cold dark matter $\omega_{\mathrm{DM}}=0$, it leads us to an
effective negative pressure $-1/3<\omega_{f}\leq0$. This way, torsion can
easily produce an effective \textquotedblleft non-particle\textquotedblright%
\ negative pressure $p_{f}$ from canonical cold dark matter $\omega
_{\mathrm{DM}}=0$.

From Eqs.~(\ref{Eq_Friedmann_Density_Seminaked})
and~(\ref{Eq_Friedmann_Pressure_Seminaked}), we may feel compelled to define a
generalized Hubble parameter $H+h$. However, it is important to remember that
our observations describe the behavior of classical particles (i.e.,
galaxies). Classical particles are sensitive only to the Riemannian piece of
the geometry and oblivious to torsion, and therefore observations measure $H$
and not $H+h$. For this reason, it is convenient to write down the
Eqs.~(\ref{Eq_Friedmann_Density_Seminaked})
and~(\ref{Eq_Friedmann_Pressure_Seminaked}) as%
\begin{align}
3H^{2}  &  =\rho_{f}+\rho_{h},\label{Eq_Friedmann_rho_h}\\
2\dot{H}+3H^{2}  &  =-\left(  p_{f}+p_{h}\right)  , \label{Eq_Friedmann_p_h}%
\end{align}
where $\rho_{h}$ and $p_{h}$ are the effective density and pressure originated
when moving all the $h\left(  t\right)  $ terms to the right-hand side of the
field equations%
\begin{align}
\rho_{h}  &  =-3\left(  h+2H\right)  h,\label{Eq_Def_rho_h}\\
p_{h}  &  =h^{2}+4Hh+2\dot{h}.
\end{align}
The total dark matter and torsion weight is described by the effective
\textquotedblleft dressed\textquotedblright\ density and pressure%
\begin{align}
\rho_{\mathrm{dressed}}  &  =\rho_{f}+\rho_{h},\\
p_{\mathrm{dressed}}  &  =p_{f}+p_{h}.
\end{align}
At this point, using the relation~(\ref{Eq_Protobaro_h}) we propose the
\textquotedblleft barotropic\textquotedblright\ Ansatz%
\begin{equation}
h=\alpha_{h}\sqrt{1+\alpha_{f}^{2}}\sqrt{\rho_{\mathrm{DM}}}=\alpha_{h}%
\sqrt{\rho_{f}},
\end{equation}
and from here the behavior of the dark matter spin tensor, and its torsion
becomes more clear. The two functions $f$ and $h$ parametrize the dark matter
spin tensor, and they create an effective \textquotedblleft torsional
dress\textquotedblright\ for $\rho_{\mathrm{DM}}$ and $p_{\mathrm{DM}}$. For
instance, a small $\rho_{\mathrm{DM}}$ may be amplified for torsion and create
a much bigger $\rho_{\mathrm{dressed}}=\rho_{f}+\rho_{h}$. Current
observations would measure the effective $\rho_{\mathrm{dressed}}$ and not the
original dark matter density $\rho_{\mathrm{DM}}$.

On the other hand, the torsional-dressed density $\rho_{\mathrm{dressed}}%
=\rho_{f}+\rho_{h}$ has more complex behavior. From
Eq.~(\ref{Eq_Friedmann_Density_Seminaked}), it is clear that $6\left(
H+h\right)  \left(  \dot{H}+\dot{h}\right)  =\dot{\rho}+\dot{\rho}_{f}$.
Replacing this in Eq.~(\ref{Eq_Friedmann_Pressure_Seminaked}) and considering
that dark matter does not interact with SM matter
Eq.~(\ref{Eq_Conservation_SM}), it is possible to prove that the two dark
matter-torsion modes, the $f$-dressed density $\rho_{f}$ and the $h$-dressed
density $\rho_{h}$, interchange energy among them%
\begin{align}
\dot{\rho}_{f}+3H\left(  \rho_{f}+p_{f}\right)   &  =-Q,\label{Eq_Q1}\\
\dot{\rho}_{h}+3H\left(  \rho_{h}+p_{h}\right)   &  =Q, \label{Eq_Q2}%
\end{align}
where%
\begin{equation}
Q=\left(  1+3\omega_{f}\right)  h\rho_{f}, \tag{46}%
\end{equation}
and therefore the effective density $\rho_{\mathrm{dressed}}=\rho_{f}+\rho
_{h}$ obeys the canonical conservation relation%
\begin{equation}
\dot{\rho}_{\mathrm{dressed}}+3H\left(  \rho_{\mathrm{dressed}}%
+p_{\mathrm{dressed}}\right)  =0, \tag{47}%
\end{equation}
with $p_{\mathrm{dressed}}$ obeying a non-trivial equation of state
$p_{\mathrm{dressed}}=p_{\mathrm{dressed}}\left(  \rho_{f},\rho_{h}\right)  $.
The next Section analyses the phenomenology of this system for some important
particular cases.

\label{Sec_Cosmo_DM_DT}

\section{Cosmological consequences of Dark Torsion}

The two torsional modes $h$ and $f$ create very distinctive phenomenology in
the context of cosmic evolution. The simplest case is $h=\alpha_{h}=0$,
leading us to a system of the canonical form%
\begin{align}
3H^{2}  &  =\rho_{f},\tag{48}\\
\dot{\rho}_{f}+3H\left(  1+\omega_{f}\right)  \rho_{f}  &  =0. \tag{49}%
\end{align}
When $h=0$, the effective density $\rho_{f}$ packs dark matter and torsion
altogether. The only difference with the standard $\mathrm{\Lambda CDM}$
torsionless case is that for cold dark matter $\omega_{\mathrm{DM}}=0$, the
effective barotropic constant $\omega_{f}=\left(  \omega_{\mathrm{DM}}%
-\alpha_{f}^{2}/3\right)  /\left(  1+\alpha_{f}^{2}\right)  $ has the allowed
range $-1/3<\omega_{f}\leq0$. Since $\rho_{f}=\left(  1+\alpha_{f}^{2}\right)
\rho_{\mathrm{DM}}$, it means that for large values of $\alpha_{f}$ a small
quantity of dark matter can get significantly amplified.

The case $h\neq0$. From Eq.~(\ref{Eq_Friedmann_Density_Seminaked}) we can obtain%

\begin{equation}
H\left(  t\right)  =\left(  \sqrt{\frac{1}{3}}\frac{\mathrm{s}_{H+h}%
}{\left\vert \alpha_{h}\right\vert }-\mathrm{s}_{h}\right)  \left\vert
h\left(  t\right)  \right\vert , \tag{50}%
\end{equation}
where we are using the shortcut notation $\mathrm{s}_{X}=\mathrm{sign}\left(
X\right)  $. Therefore the $\rho_{h}$ density corresponds to%

\begin{equation}
\rho_{h}=3\left(  \left\vert \alpha_{h}\right\vert -\frac{2}{\sqrt{3}%
}\mathrm{s}_{h}\mathrm{s}_{H+h}\right)  \frac{h^{2}}{\left\vert \alpha
_{h}\right\vert }>0, \tag{51}%
\end{equation}
but
\begin{align}
\mathrm{s}_{h}  &  =-1\text{ \ },\text{ \ }\mathrm{s}_{H+h}=1\Longrightarrow
H\left(  t\right)  =\left(  \sqrt{\frac{1}{3}}\frac{\mathrm{1}}{\left\vert
\alpha_{h}\right\vert }+\mathrm{1}\right)  \left\vert h\left(  t\right)
\right\vert \rightarrow\frac{\left\vert h\left(  t\right)  \right\vert
}{H\left(  t\right)  }\text{ }<1\text{\ },\text{ \ }\rho_{h}=3\left(
\left\vert \alpha_{h}\right\vert +\frac{2}{\sqrt{3}}\right)  \frac{h^{2}%
}{\left\vert \alpha_{h}\right\vert },\tag{52}\\
\mathrm{s}_{h}  &  =-1\text{ \ },\text{ \ }\mathrm{s}_{H+h}=-1\Longrightarrow
H\left(  t\right)  =\left(  1-\sqrt{\frac{1}{3}}\frac{\mathrm{1}}{\left\vert
\alpha_{h}\right\vert }\right)  \left\vert h\left(  t\right)  \right\vert
\Longrightarrow\left\vert \alpha_{h}\right\vert >\sqrt{\frac{1}{3}}%
\rightarrow\rho_{h}=3\left(  \left\vert \alpha_{h}\right\vert -\frac{2}%
{\sqrt{3}}\right)  \frac{h^{2}}{\left\vert \alpha_{h}\right\vert },\tag{53}\\
\mathrm{s}_{h}  &  =1\text{ \ },\text{ \ }\mathrm{s}_{H+h}=1\Longrightarrow
H\left(  t\right)  =\left(  \sqrt{\frac{1}{3}}\frac{\mathrm{1}}{\left\vert
\alpha_{h}\right\vert }-\mathrm{1}\right)  \left\vert h\left(  t\right)
\right\vert \Longrightarrow\left\vert \alpha_{h}\right\vert <\sqrt{\frac{1}%
{3}}\rightarrow\rho_{h}=3\left(  \left\vert \alpha_{h}\right\vert -\frac
{2}{\sqrt{3}}\right)  \frac{h^{2}}{\left\vert \alpha_{h}\right\vert }<0,
\tag{54}%
\end{align}
and the last two cases lead to $\left\vert h\left(  t\right)  \right\vert
/H\left(  t\right)  >1$. Additionally we note that the case $\mathrm{s}%
_{h}=-1$, $\mathrm{s}_{H+h}=1$ leads directly to $\rho_{h}>0$ and we are well
with the weak energy condition. But, is it reasonable to expect $\left\vert
h\left(  t\right)  \right\vert /H\left(  t\right)  <1$ or $\left\vert h\left(
t\right)  \right\vert /H\left(  t\right)  <<1$ during the cosmic evolution?
This is an open question.

Now, after replacing $H$ and $h=\alpha_{h}\sqrt{\rho_{f}}$ into $\dot{\rho
}_{f}+3H\left(  1+\omega_{f}\right)  \rho_{f}=-\left(  1+3\omega_{f}\right)
h\rho_{f}$, we obtain the following solution for $h$, with $\mathrm{s}_{h}=-1$
and\ $\mathrm{s}_{H+h}=1$,%

\begin{equation}
\frac{h\left(  t\right)  }{h\left(  t_{0}\right)  }=\left[  1+\Delta\left(
t-t_{0}\right)  \right]  ^{-1}, \tag{55}%
\end{equation}
where $t_{0}$ is today and%

\begin{equation}
\Delta=\left[  \left\vert \alpha_{h}\right\vert -\frac{\sqrt{3}}{2}\left(
1+\omega_{f}\right)  \right]  \frac{\left\vert h\left(  t_{0}\right)
\right\vert }{\left\vert \alpha_{h}\right\vert }. \tag{56}%
\end{equation}
Recalling that $\omega_{\mathrm{DM}}=0\rightarrow\omega_{f}=-\alpha_{f}%
^{2}/3\left(  1+\alpha_{f}^{2}\right)  $, we write%

\begin{equation}
\Delta=\left[  \left\vert \alpha_{h}\right\vert -\frac{\sqrt{3}}{2}\left(
\frac{1+2\alpha_{f}^{2}/3}{1+\alpha_{f}^{2}}\right)  \right]  \frac{\left\vert
h\left(  t_{0}\right)  \right\vert }{\left\vert \alpha_{h}\right\vert },
\tag{57}%
\end{equation}
so that%

\begin{align}
\mathrm{s}_{\Delta}  &  =1\Longrightarrow\Delta>0\rightarrow\text{standard
scheme},\tag{58}\\
\mathrm{s}_{\Delta}  &  =-1\Longrightarrow\Delta<0\rightarrow\text{phantom
evolution!} \tag{59}%
\end{align}
But%

\begin{equation}
\Delta<0\longleftrightarrow\left\vert \alpha_{h}\right\vert <\frac{\sqrt{3}%
}{2}\left(  \frac{1+2\alpha_{f}^{2}/3}{1+\alpha_{f}^{2}}\right)  <1, \tag{60}%
\end{equation}
and so%

\begin{equation}
h\left(  t\right)  =\frac{h\left(  t_{0}\right)  }{\left\vert \Delta
\right\vert }\left(  t_{s}-t\right)  ^{-1}\text{ \ },\text{ \ }t_{s}%
=t_{0}+\frac{1}{\left\vert \Delta\right\vert }, \tag{61}%
\end{equation}
and all the components explode at $t_{\mathrm{s}}$: the Hubble parameter, the
densities $\rho_{\mathrm{DM}}$, $\rho_{f}$, $\rho_{\mathrm{dressed}}$ and $Q$.
This a Big Rip singularity.

The $Q$-function becomes%
\begin{equation}
Q\left(  t\right)  =-\frac{1}{\left\vert a_{h}\right\vert ^{2}\left(
1+\alpha_{f}^{2}\right)  }\left\vert h\right\vert ^{3}, \tag{62}%
\end{equation}
\ \ \ \ and so, there is energy transference from $\rho_{h}$ to $\rho_{f}$.

\section{Thermodynamics}

We inspect two thermodynamics aspects in presence of torsion, adiabaticity and
dark matter temperature.\ We start with the Gibb's relation%

\begin{equation}
T\mathrm{d}S=\mathrm{d}\left(  \frac{\rho_{\mathrm{DM}}}{n}\right)
+p_{_{\mathrm{DM}}}\mathrm{d}\left(  \frac{1}{n}\right)  , \tag{63}%
\end{equation}
implying%
\begin{equation}
nT\frac{\mathrm{d}S}{\mathrm{d}t}=-\left(  \rho_{_{\mathrm{DM}}}%
+p_{_{\mathrm{DM}}}\right)  \frac{\dot{n}}{n}+\dot{\rho}_{\mathrm{DM}},
\tag{64}%
\end{equation}
where $T$ is the temperature, $S$ the entropy and $n$ the number particle
density. Using the integrability condition ($S$ is a function of state) we
have $\partial^{2}S/\partial T\partial n=\partial^{2}S/\partial n\partial T$
and therefore%

\begin{equation}
n\frac{\partial T}{\partial n}+\left(  p_{\mathrm{DM}}+\rho_{\mathrm{DM}%
}\right)  \frac{\partial T}{\partial\rho_{\mathrm{DM}}}=T\frac{\partial
p_{_{\mathrm{DM}}}}{\partial\rho_{\mathrm{DM}}}. \tag{65}%
\end{equation}
Since $\rho_{f}=\left(  1+\alpha_{f}^{2}\right)  \rho_{\mathrm{DM}}$ and
$\rho_{f}$ satisfies Eq.~(\ref{Eq_Q1}), we have that the bare dark matter
density obeys the \textquotedblleft conservation\textquotedblright\ law%

\begin{equation}
\dot{\rho}_{\mathrm{DM}}+3H\left(  1+\omega_{f}\right)  \rho_{\mathrm{DM}%
}=-\frac{Q}{1+\alpha_{f}^{2}}, \tag{66}%
\end{equation}
and making the hypothesis of $\dot{n}+3Hn=0$ (conservation of the number of
dark matter particles) we have that
\begin{equation}
nT\frac{\mathrm{d}S}{\mathrm{d}t}=3H\left(  p_{_{\mathrm{DM}}}-\omega_{f}%
\rho_{\mathrm{DM}}\right)  -\frac{Q}{1+\alpha_{f}^{2}}. \tag{67}%
\end{equation}
In the case of cold dark matter $\omega_{\mathrm{DM}}=0$ it implies that%

\begin{equation}
nT\frac{\mathrm{d}S}{\mathrm{d}t}=\frac{1}{1+\alpha_{f}^{2}}\left(
H\alpha_{f}^{2}\rho_{\mathrm{DM}}-Q\right)  , \tag{68}%
\end{equation}
and there is not adiabaticity. Given that $Q<0$, then $dS/dt>0$ and the second
law of thermodynamics is guaranteed.

As we know, $\Lambda CDM$~\cite{Ref-PLANCK} has been quite successful in
describing the current state of cosmic evolution even when it is not free of
problems. As a consequence of this, the idea of dark energy emerged as a more
physical alternative to $\Lambda$ and, moreover, dark matter-dark energy
interaction is a fact does not ruled out by observation~\cite{Ref-B.Wang}. In
this interaction framework, the non-adiabaticity is manifest~\cite{Ref-Victor}%
. So, we conjecture that torsion is cause of non-adiabaticity!

Using the integrability condition,\ the temperature can be obtained from%

\begin{equation}
\frac{\dot{T}}{T}=-3H\omega_{\mathrm{DM}}\left(  1+\frac{\left(
1+\omega_{\mathrm{DM}}\right)  \left(  1+\alpha_{f}^{2}\right)  }%
{1+\omega_{\mathrm{DM}}+2\alpha_{f}^{2}/3+Q/3H\rho_{\mathrm{DM}}}\right)
^{-1}, \tag{69}%
\end{equation}
and it is clear that $\omega_{\mathrm{DM}}=0\Longrightarrow T=\mathrm{const}%
$., consistent with "orthodoxy" which tells us that the dark matter
temperature is constant during the cosmic evolution. Thus, torsion does not
affect the dark matter temperature.

\section{Final remarks}

\label{Sec_TheEnd}

We have found a phantom scheme originated after considering torsion
\textquotedblleft coupled\textquotedblright\ to the dark matter. As a
consequence of this, we would not need dark energy (phantom dark energy
described by $\omega_{ph}<-1$) in order to explain such late evolution. This
is an alternative that seeks to explain the phantom scheme, not ruled out by
the current observational information.

Another interesting fact that we have found is an interaction scheme between
the torsional components $h$ and $f$. The significance of this interaction is
not clear to us yet. If this interaction is such, is there any way to detect
any observational consequence from this? If torsion effects cannot be detected
with the current observations, perhaps it can be done in the future if the
polarization of gravitational waves is measured and from this observational
fact we can also have indications of $\left\vert h\left(  t\right)
\right\vert /H\left(  t\right)  $. This is an interesting conjecture to explore.

From the thermodynamic point of view, the cosmic evolution turns out to be
non-adiabatic ($\mathrm{\Lambda CDM}$ is an adiabatic scheme) and since $h<0$,
the second law of thermodynamics is guaranteed. The interesting thing is also
that the dark matter temperature, even in the presence of torsion, remains
constant through the cosmic evolution.

Finally, and as we have already said, future observations could shed some
light on the role, if any, of torsion in the cosmic evolution.

\begin{acknowledgement}
FI acknowledges financial support from the Chilean government through FONDECYT
grant 1180681 of the Government of Chile.
\end{acknowledgement}

\end{document}